\documentclass[10pt]{article}

\usepackage{amsmath}
\usepackage{amssymb}

\usepackage{graphicx}

\usepackage{cite}

\usepackage{color} 

\usepackage[labelfont=bf,labelsep=period,justification=raggedright]{caption}

\makeatletter
\renewcommand{\@biblabel}[1]{\quad#1.}
\makeatother

\date{}

\begin{document}

\begin{flushleft}
{\Large
\textbf{These are not the k-mers you are looking for: efficient
online k-mer counting using a probabilistic data structure}
}
\\
Qingpeng Zhang$^{1}$, 
Jason Pell$^{1}$,
Rosangela Canino-Koning$^{1}$,
Adina Chuang Howe$^{2,3}$,
C. Titus Brown$^{1,2\ast}$
\\
\bf{1} Department of Computer Science and Engineering, Michigan State University,
East Lansing, MI, USA
\\
\bf{2} Department of Microbiology and Molecular Genetics, Michigan State University,
East Lansing, MI, USA
\\
\bf{3} Department of Plant, Soil, and Microbial Sciences, Michigan State University, 
East Lansing, MI, USA
\\
$\ast$ E-mail: ctb@msu.edu
\end{flushleft}





\section*{Abstract}

K-mer abundance analysis is widely used for many purposes in nucleotide sequence
analysis, including data preprocessing for de novo assembly, repeat
detection, and sequencing coverage estimation.  We present the khmer
software package for fast and memory efficient {\em online} counting
of k-mers in sequencing data sets. Unlike previous methods based on
data structures such as hash tables, suffix arrays, and trie
structures, khmer relies entirely on a simple probabilistic data
structure, a Count-Min Sketch.  The Count-Min Sketch permits online
updating and retrieval of k-mer counts in memory which is necessary to
support online k-mer analysis algorithms.  On sparse data sets this
data structure is considerably more memory efficient than any exact
data structure.  In exchange, the use of a Count-Min Sketch introduces
a systematic overcount for k-mers; moreover, only the counts, and not
the k-mers, are stored.  Here we analyze the speed, the memory usage,
and the miscount rate of khmer for generating k-mer frequency
distributions and retrieving k-mer counts for individual k-mers.  We
also compare the performance of khmer to several other k-mer counting
packages, including Tallymer, Jellyfish, BFCounter, DSK, KMC, Turtle
 and KAnalyze.
Finally, we examine
the effectiveness of profiling sequencing error, k-mer abundance
trimming, and digital normalization of reads in the context of high
khmer false positive rates. khmer is implemented in C++ wrapped in a Python
interface, offers a tested and robust API, and is freely available
under the BSD license at github.com/ged-lab/khmer.


\section*{Introduction}

The goal of k-mer counting is to determine the number of occurrences
for each fixed-length word of length k in a DNA data set
\cite{Marcais2011}. Efficient k-mer counting plays an important role
in many bioinformatics approaches, including data preprocessing for de
novo assembly, repeat detection, and sequencing coverage estimation
\cite{Kurtz2008}.


Short-read shotgun sequencing data is both relatively sparse in k-mers
and contains many erroneous k-mers.  For typical values of k such as
32 these data sets are sparse, as only a small fraction of the total
possible number of k-mers ($4^{32}$) are actually present in any
genome or read data sets derived from the genome.  The high error rate
(e.g. Illumina has a ~0.1-1\% per-base error rate
\cite{pubmed19997069}) generates many unique k-mers.  As the total
number of generated reads increases, the total number of errors grows
with it linearly. This leads to data sets where the erroneous k-mers
vastly outnumber the true k-mers \cite{Conway2011}.  Tracking and
counting the resulting large number of k-mers, most of which are
erroneous, has become an unavoidable and challenging task in sequence
analysis \cite{Minoche2011}.

A variety of k-mer counting approaches, and standalone software
packages implementing them, have emerged in recent years; this
includes Tallymer, Jellyfish, BFCounter, DSK, KMC, Turtle and KAnalyze
\cite{Kurtz2008, Marcais2011, Melsted2011, Rizk2013, Deorowicz2013,
  Roy2014, Audano2014}.

These approaches and implementations each offer different algorithmic
trade-offs and enable a non-overlapping set of functionality.
Tallymer uses a suffix tree to store k-mer counts in memory and on
disk \cite{Kurtz2008}.  Jellyfish stores k-mer counts in in-memory
hash tables, and makes use of disk storage to scale to larger data
sets \cite{Marcais2011}.  BFCounter uses a Bloom filter as a
pre-filter to avoid counting unique k-mers, and is the first published
probabilistic approach to k-mer counting \cite{Melsted2011}.  DSK
adopts an
approach to k-mer counting that enables time- and
memory-efficient k-mer counting with an explicit trade-off between
disk and memory usage \cite{Rizk2013}.  KMC and KAnalyze rely
primarily on fast and inexpensive disk access to count k-mers in low
memory \cite{Deorowicz2013,Audano2014}.  Turtle provides several
different containers that offer different false positive and false
negative tradeoffs when counting k-mers \cite{Roy2014}.

Our motivation for exploring efficient k-mer counting comes from our
work with metagenomic data, where we routinely encounter data sets
that contain $300 \times 10^9$ bases of DNA and over 50 billion
distinct k-mers \cite{Howe2012}.  To efficiently filter, partition,
and assemble these data, we need to store counts for each of these
k-mers in main memory, and query and update them in realtime --- a set
of functionality not readily offered by current packages.  Moreover,
we wish to enable the use of cloud and desktop computers, which may
have poor I/O performance or limited memory. These needs have dictated
our exploration of efficient in-memory k-mer counting techniques.


Below, we describe an implementation of a simple probabilistic data
structure for k-mer counting. This data structure is based on a
Count-Min Sketch \cite{Cormode2005}, a generalized probabilistic data structure for
storing the frequency distributions of distinct elements.
Our implementation extends an earlier
implementation of a Bloom filter \cite{Bloom70}, which has been previously used 
in bioinformatics applications, such as
sequence matching \cite{DBLP:conf/padl/MaldeO09}, k-mer counting \cite{Melsted2011}, 
and de Bruijn graph storage and traversal \cite{Pell2012,Jones:2012aa}.
Many other variations of Bloom filters have been proposed \cite{BroderM03}, including
counting Bloom filters \cite{Fan:2000:SCS:343571.343572}, 
multistage filters \cite{DBLP:conf/sigcomm/EstanV02}, 
and spectral Bloom filters \cite{DBLP:conf/sigmod/CohenM03}, which 
are related to the Count-Min Sketch and our khmer implementation.

Probabilistic approaches can be particularly memory efficient for
certain problems, with memory usage significantly lower than any exact
data structure \cite{Pell2012}.  However, their use introduces set
membership or counting false positives, which have effects that must
be analyzed in the context of specific problems.  Moreover, unlike
existing techniques, the Count-Min Sketch stores only counts; k-mers
must be retrieved from the original data set.  In exchange, the low
memory footprint enabled by this probabilistic approach enables online
updating and retrieval of k-mer counts entirely in memory, which in
turn supports streaming applications such as digital normalization
\cite{Brown2012}.

We use the Amazon cloud to compare time, memory, and disk usage of our
k-mer counting implementation with that of other k-mer counting software packages, 
for two problems. First, we generate a k-mer abundance
distribution for large data sets; and second, we query many individual
k-mer counts at random from a previously constructed k-mer count
database.  We show that khmer is competitive in speed, memory, and
disk usage for these problems.  We also analyze the effects of
counting error on calculations of the k-mer count in sequencing data
sets, and in particular on metagenomic data sets.  Finally, we discuss
khmer's miscount performance in the context of two specific
applications: low-abundance k-mer trimming of reads, and digital
normalization.

The khmer software \cite{khmer} is implemented in C++ in a Python
wrapper, enabling flexible use and reuse by users with a wide range of
computational expertise.  The software package is freely available for
academic and commercial use and redistribution under the BSD license
at github.com/ged-lab/khmer/.  khmer comes with substantial
documentation and many tutorials, and contains extensive unit tests.
Moreover, we have built several applications on top of khmer,
including memory-efficient de Bruijn graph partitioning
\cite{Pell2012} and lossy compression of short-read data sets for
assembly \cite{Brown2012}.

\section*{Results}

\subsection*{Implementing a Count-Min Sketch for k-mers}

The two basic operations supported by khmer are {\tt
  increment\_count(kmer) } and {\tt c = get\_count(kmer). } Both
operate on the data structure in memory, such that neither
incrementing a count nor retrieving a count involves disk access.

The implementation details are similar to those of the Bloom filter in
\cite{Pell2012}, but with the use of 8 bit counters instead of 1 bit
counters.  Briefly, Z hash tables are allocated, each with a different
size of approximately H bytes ($H_1, H_2, ..., H_Z$); the sum of these
hash table sizes must fit within available main memory.  To increment
the count for a particular k-mer, a single hash is computed for the
k-mer, and the modulus of that hash with each hash table's size H
gives the location for each hash table; the associated count in each
hash table is then incremented by 1.  We use different sizes for each
hash table so as to vary the hash function.  Even if two k-mers have
the same modulus in one hash table (a collision), they are unlikely to
collide in the other hash tables.  To retrieve the count for a k-mer,
the same hash is computed and the minimum count across all hash tables
is computed. While different in implementation detail from the
standard Bloom filter, which uses a single hash table with many
hash functions, the performance details are identical \cite{Pell2012}.
One particularly important feature of the Count-Min Sketch is that the
counting error is {\em one-sided} \cite{Cormode2005}.  Because counts
are only incremented, collisions result in inflated miscounts; if
there is no collision for a particular k-mer, the count is correct.

An additional benefit of the Count-Min Sketch is that it is extremely
easy to implement correctly, needing only about 3 dozen lines of C++
code for a simple threadsafe implementation.  (We have
described how khmer scales with multiple threads in
\cite{McDonald2013}.)

To determine the expected false positive rate --- the average frequency with
which a given k-mer count will be incorrect when retrieved --- we can
look at the hash table load. Suppose N distinct k-mers have been counted
using Z hash tables, each with size H.  The probability that no
collisions happened in a specific entry in one hash table is
$(1-1/H)^{N}$, or approximately $e^{-N/H}$. The individual collision
rate in one hash table is then $\approx 1-e^{-N/H}$. The total
collision rate, which is the probability that a collision occurred in
each entry where a k-mer maps across all Z hash tables, is $\approx
(1-e^{-N/H})^{Z}$, which is also the expected false positive rate.

While the false positive rate can easily be calculated from the hash table
load, the average {\em miscount} --- the degree to which the measured
count differs from the true count --- depends on the k-mer frequency
distribution, which must be determined empirically.  We analyze the
effects of this below.

\subsection*{Choosing number and size of hash tables used for k-mer counting}

The false positive rate depends on the number of distinct k-mers $N$,
the number of hash tables $Z$, and the size of the hash tables $H$: $f
\approx (1-e^{-N/H})^{Z}$, with an associated memory usage of $M = H
Z$.  We face two common scenarios: one in which we have a fixed number
of k-mers $N$ and fixed memory $M$ and we want to calculate the
optimal number of hash tables $Z$; and one in which we have a desired
maximum false positive rate $f$ and a fixed number of k-mers $N$, and
we want to calculate the minimum memory usage required to achieve $f$.

For fixed memory $M$ and number of distinct k-mers $N$, the optimal
number of hash tables can be found by minimizing $f$; taking the
derivative, $df/dZ$, with $f \approx \exp(Z \log(1-e^{-ZN/M}))$ and solving
for 0, we find that $f$ is minimized when $Z=\log(2)*(M/N)$ (see
\cite{broder2004network} for details).

Given a desired false positive rate $f$ and a fixed number of k-mers
$N$, the optimal memory usage can be calculated as follows.  First,
the optimal number of hash tables is determined by the expected false
positive rate alone: $Z = \log_{0.5}f$.  Using this $Z$, the minimum
average hash table size $H$ necessary to achieve $f$ can be calculated
as $H = (\log_{0.6185}(f)\times N)/Z$ (see
\cite{broder2004network} for details).

A remaining problem is that the number of distinct k-mers $N$ is
typically not known.  However, memory- and time-efficient algorithms
for calculating $N$ do exist and we plan to implement this in
khmer in the future \cite{flajolet2008hyperloglog}.

\subsection*{khmer efficiently calculates k-mer abundance histograms}

\begin{figure}[!ht]
\centerline{\includegraphics[width=5in]{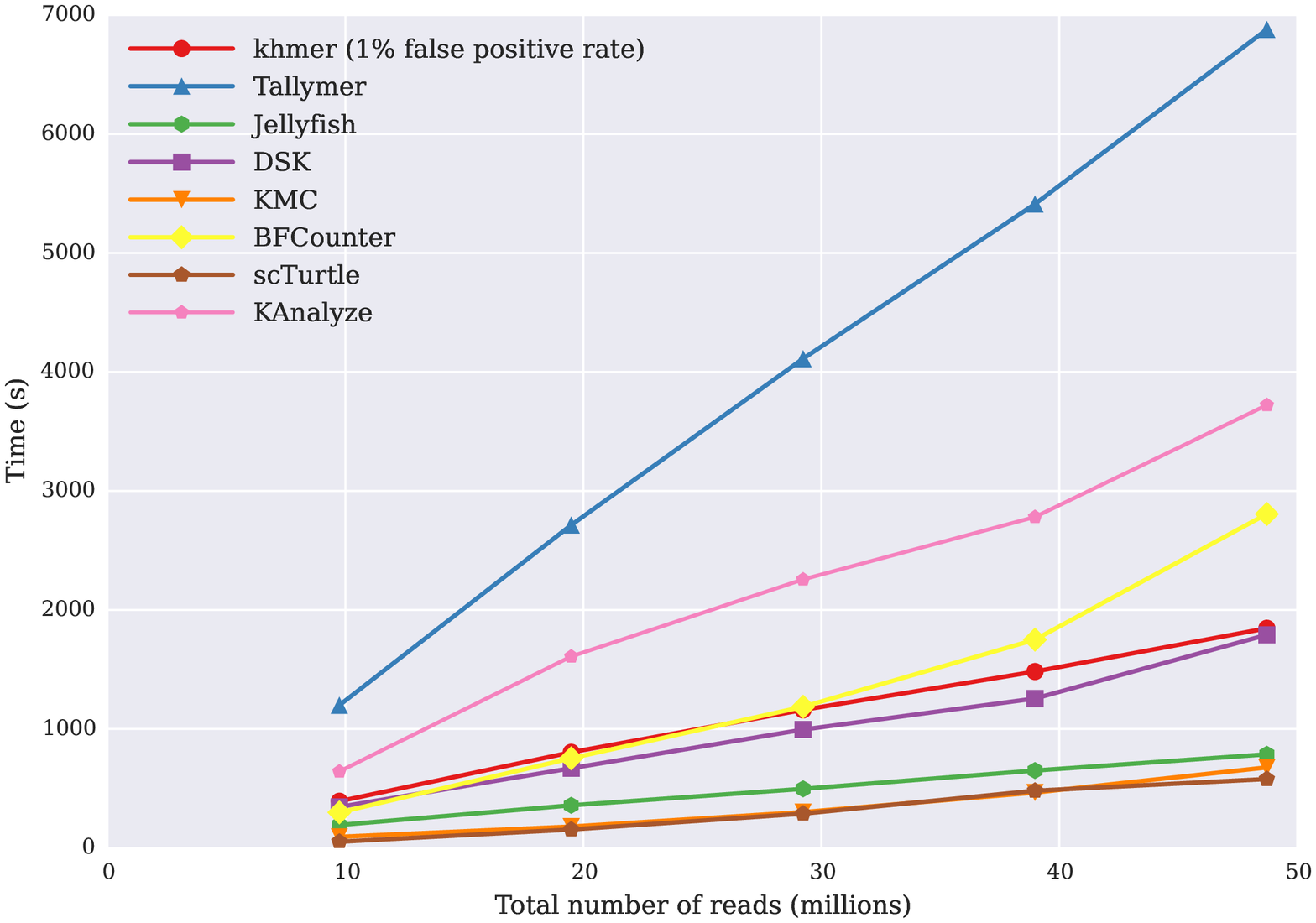}}

\caption{\bf Comparison of the time it takes for k-mer counting tools
  to calculate k-mer abundance histograms, with time (y axis, in
  seconds) against data set size (in number of reads, x axis).  
  All programs executed in time approximately linear with
  the number of input reads.}

\label{fig:cmp_time}
\end{figure}

\begin{figure}[!ht]
\centerline{\includegraphics[width=5in]{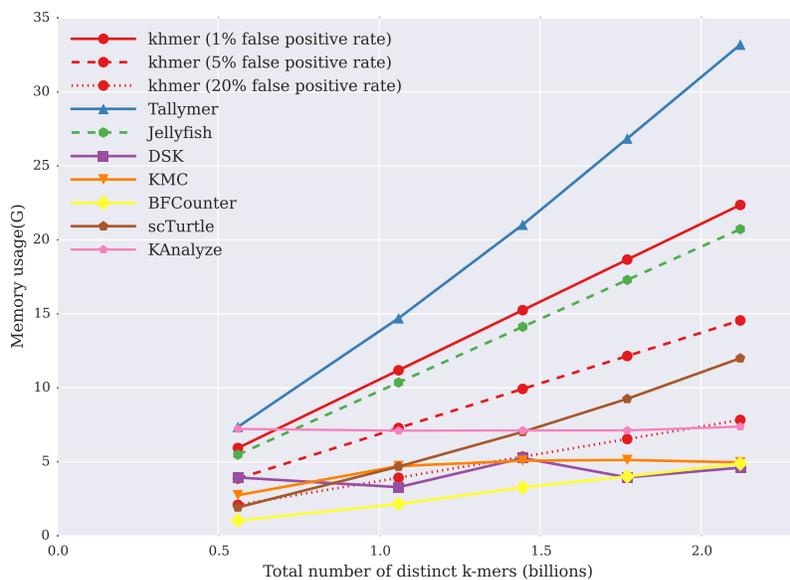}}

\caption{\bf Memory usage of k-mer counting tools when calculating
  k-mer abundance histograms, with maximum resident program size (y
  axis, in GB) plotted against the total number of distinct k-mers in
  the data set (x axis, billions of k-mers). }

\label{fig:cmp_memory}
\end{figure}

We measured time and memory required to calculate k-mer abundance
histograms in five soil metagenomic read data sets using khmer,
Tallymer, Jellyfish, DSK, KMC, Turtle, and KAnalyze (Table
\ref{table:datasets}; Figures \ref{fig:cmp_time} and
\ref{fig:cmp_memory}).  We chose to benchmark abundance histograms
because this functionality is common to all the software packages, and
is a common analysis approach for determining assembly parameters
\cite{Chikhi:2014aa}.  We applied each package to increasingly large
subsets of a 50m read soil metagenome data set \cite{Howe2012}. For
the BFCounter, KMC, Turtle and KAnalyze packages, which do not
generate k-mer abundance distribution directly, we output the
frequency of each k-mer to a file but do no further analysis.

Figure \ref{fig:cmp_time} shows that the time usage of the khmer
approach is comparable to DSK and BFCounter, and, as expected,
increases linearly with data set size. Tallymer is the slowest of the
four tools in this testing, while KMC, Turtle, and Jellyfish are
the fastest.

From Figure \ref{fig:cmp_memory}, we see that the memory usage of
Jellyfish, Tallymer, BFCounter, and Turtle increases linearly with
data set size. Tallymer uses more memory than Jellyfish generally,
while BFCounter and Turtle have considerably lower memory usage.
DSK, KMC, and KAnalyze use constant memory across the data sets, but
at the cost of more limited functionality (discussed below).


The memory usage of khmer also increases linearly with data set size
as long as we hold the false positive rate constant.  However, the
memory usage of khmer varies substantially with the desired false
positive rate: we can decrease the memory usage by increasing the
false positive rate as shown in Figure \ref{fig:cmp_memory}.  We also
see that with a low false positive of 1\%, the memory usage is
competitive with Tallymer and Jellyfish; with a higher 5\% false
positive rate, the memory usage is lower than all but the disk-based
DSK; with an false positive rate as high as 20\%, the memory usage is
further lower, close to DSK, KAnalyze, and KMC.

\begin{figure}[!ht]
\centerline{\includegraphics[width=5in]{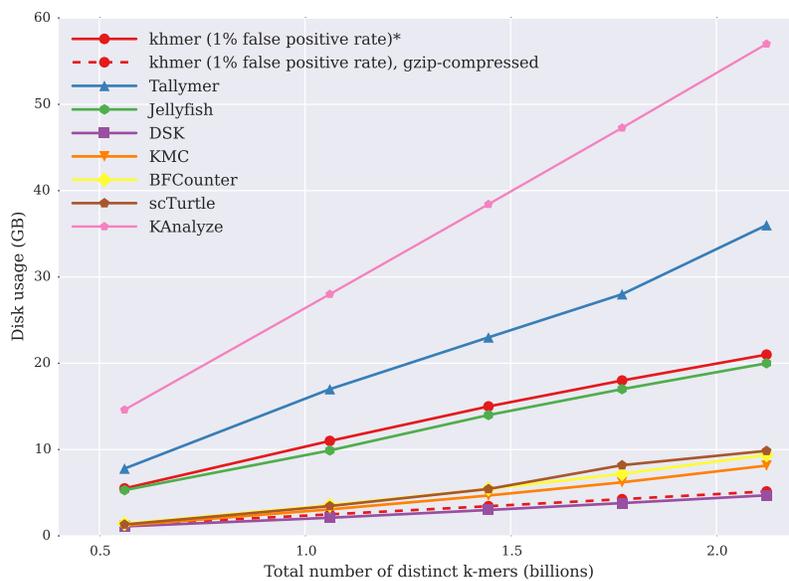}}

\caption{\bf Disk storage usage of different k-mer counting tools to
  calculate k-mer abundance histograms in GB (y axis), plotted against
  the number of distinct k-mers in the data set (x axis).  $^*$Note
  that khmer does not use the disk during counting or retrieval,
  although its hash tables can be saved for reuse.}

\label{fig:cmp_disk}
\end{figure}

We also measured disk usage during counting.  Figure
\ref{fig:cmp_disk} shows that the disk usage also increases linearly
with the number of k-mers in the data set.  For a high-diversity
metagenomic data set of 5 GB, the disk usage of both Jellyfish and
Tallymer is around 30 GB.  khmer counts k-mers entirely in working
memory and does not rely on any on-disk storage to store or retrieve
k-mer counts, although for practicality the hash tables can be saved
for later reuse; the uncompressed disk usage for khmer in Figure
\ref{fig:cmp_disk} is the same as its memory.  At the expense of more
time, khmer supports saving and loading gzip-compressed hash tables,
which are competitive in size to DSK's on-disk database (Figure 3,
dashed line).

\subsection*{khmer accesses k-mer counts efficiently}

\begin{figure}[!ht]
\centerline{\includegraphics[width=5in]{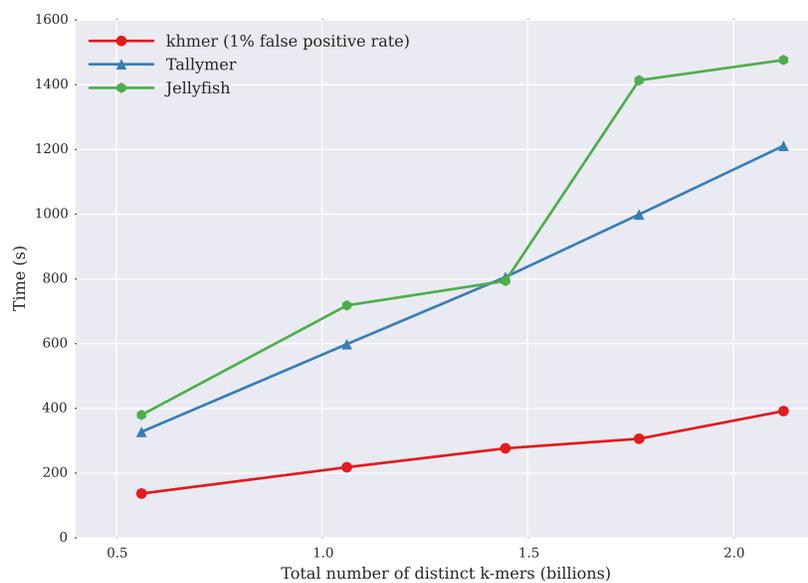}}
\caption{\bf Time for several k-mer counting tools to retrieve the
  counts of 9.7m randomly chosen k-mers (y axis), plotted against the
  number of distinct k-mers in the data set being queried (x axis).
  BFCounter, DSK, Turtle, KAnalyze, and KMC do not support this functionality.}
\label{fig:cmp_count}
\end{figure}

We measured the time it took to access 9.7m 22-mers across five
different data sets after the initial databases had been built (Figure
\ref{fig:cmp_count}).  Note that Tallymer, Jellyfish, and khmer all
support random access to k-mer counts, while BFCounter, DSK, KMC, Turtle and KAnalyze 
do not. Here, khmer
performed well, dramatically outperforming Jellyfish and Tallymer.  In
all three cases, system time dominated the overall time required to
retrieve k-mers, suggesting that the primary reason for the increase
in retrieval time was due to the increased size of the database on the
disk (data not shown).  In particular, khmer is independent of the
size of the database in retrieval time once the hash tables are loaded
into memory.

\subsection*{The measured counting error is low on short-read data}

Due to the use of Count-Min Sketch and its lack of collision tracking,
khmer will report some incorrect counts for k-mers; these counts are
always higher than the true counts, up to the bound of 255 (a limit
imposed by our use of 8-bit counters). The frequency with which
incorrect counts are reported can be estimated from the hash table
load.  However, the expected {\em miscount} --- the difference between
the true k-mer frequency and the reported k-mer frequency --- cannot
be calculated without knowing the distribution of k-mer abundances; in
general, the average miscount will be small if the data is
left-skewed.  As noted by Melsted and Pritchard, a large number of
k-mers in short-read data are low-abundance, leading to precisely the
skew that would yield low miscounts \cite{Melsted2011}.  Here we use
both real and simulated data sets to evaluate the counting performance
in practice.

\begin{figure}[!ht]
\centerline{\includegraphics[width=5in]{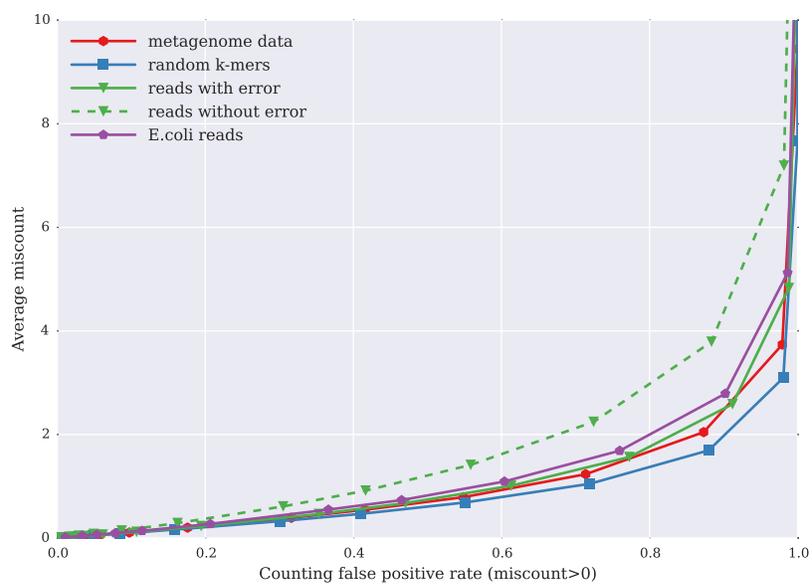}}
\caption{\bf Relation between average miscount --- amount by which
the count for k-mers is incorrect --- on the y axis, plotted against
false positive rate (x axis), for five data sets.  The five data
sets were chosen to have the same total number of distinct k-mers: one
metagenome data set; a set of randomly generated k-mers; a set
of reads, chosen with 3x coverage and 1\% error, from a randomly generated
genome; a simulated set of error-free reads (3x) chosen from a randomly
generated genome and a set of {\em E. coli} reads.}
\label{fig:average_offset_vs_fpr}
\end{figure}

Figure \ref{fig:average_offset_vs_fpr} shows the relationship between
average miscount and counting false positive rate for five different test data
sets with similar numbers of distinct k-mers: one metagenome data set;
a simulated set of random k-mers; a simulated set of reads, chosen
with 3x coverage and 1\% error; a simulated set of reads (3x) with no
error; and a set of {\em E. coli} reads (Table
\ref{table:random_data}).  Even when the counting false positive rate is as
high as 0.9 --- where 90\% of k-mers have an incorrect count --- the
average miscount is still below 4.

\begin{figure}[!ht]
\centerline{\includegraphics[width=5in]{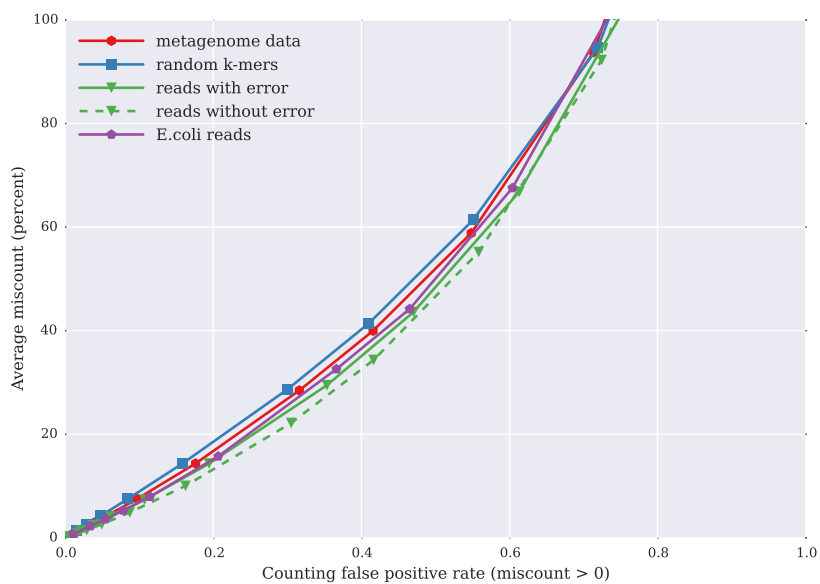}}
\caption{\bf Relation between percent miscount --- amount by which the
  count for k-mers is incorrect relative to its true count --- on the
  y axis, plotted against false positive rate (x axis), for five data
  sets.  The five data sets are the same as in Figure
  \ref{fig:average_offset_vs_fpr}.}
\label{fig:percent_offset_vs_fpr}
\end{figure}

We separately analyzed the average {\em percentage} miscount between
true and false k-mers; e.g. an miscount of 4 for a k-mer whose true
count is 1 would be 400\%.  Figure \ref{fig:percent_offset_vs_fpr}
shows the relationship between average miscount and counting false
positive rate for the same five data sets as in Figure
\ref{fig:average_offset_vs_fpr}.  For a false positive rate of 0.1 (10\% of
k-mer counts are incorrect), the average percentage miscount is less
than 10\% for all five data sets; this will of course generally be
true, because the average miscount is bounded by the product of the
false positive rate with k-mer abundance.

We see here that for a fixed false positive rate, the simulated reads
without error have the highest average miscount, and the randomly
generated k-mers have the lowest average miscount.  This is because
these two abundance distributions have the least and most left-skew,
respectively: the simulated reads without error have no abundance-1
k-mers, while the randomly generated k-mers are entirely low
abundance.

\subsection*{Sequencing error profiles can be measured with k-mer abundance
profiles}

One specific use for khmer is detecting random sequencing errors by
looking at the k-mer abundance distribution within reads
\cite{Medvedev2011}.  This approach, known also as ``k-mer spectral
analysis'', was first proposed in by \cite{Pevzner2001} and further
developed in \cite{Li2003}.  The essential idea is that low-abundance
k-mers contained in a high-coverage data set typically represent
random sequencing errors.

A variety of read trimming and error correcting tools use k-mer
counting to reduce the error content of the read data set, independent
of quality scores or reference genomes \cite{Kelley2010}.  This is an
application where the counting error of the Count-Min Sketch approach
used by khmer may be particularly tolerable: it will never falsely
call a high-abundance k-mer as low-abundance because khmer never
underestimates counts.

\begin{figure}[!ht]
\centerline{\includegraphics[width=5in]{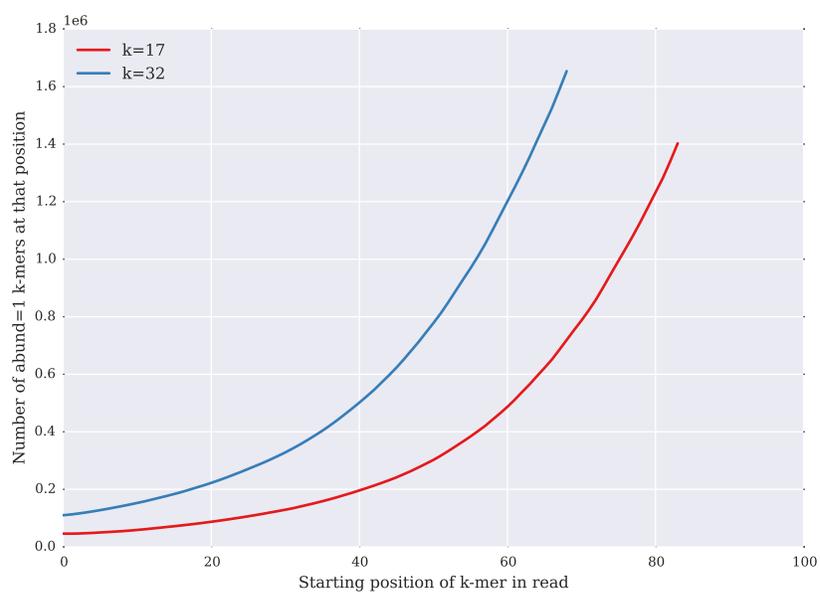}}
\caption{\bf Number of unique k-mers (y axis) by starting position
  within read (x axis) in an untrimmed {\em E. coli} 100-bp Illumina
  shotgun data set, for k=17 and k=32.  The increasing numbers of
  unique k-mers are a sign of the increasing sequencing error towards
  the 3' end of reads.  Note that there are only 69 starting positions
  for 32-mers in a 100 base read.}
\label{fig:perc_unique_pos}
\end{figure}

In Figure \ref{fig:perc_unique_pos}, we use khmer to examine the
sequencing error pattern of a 5m-read subset of an Illumina reads data
set from single-colony sequencing of {\em E. coli}
\cite{pubmed21926975}.  The high rate of occurrence of unique k-mers
close to the 3' end of reads is due to the increased sequencing error
rate at the 3' end of reads.

\subsection*{khmer can be applied iteratively to read trimming}

We next evaluated the effect of false-positive induced miscounts on
read trimming, in which reads are truncated at the first low-abundance
k-mer.  Because the Count-Min Sketch never undercounts k-mers, reads
will never be erroneously trimmed at truly high-abundance k-mers;
however, reads may not be trimmed correctly when miscounts inflate the
count of low-abundance k-mers.  In cases where many errors remain,
read trimming can potentially be applied multiple times, with each
round reducing the total number of k-mers and hence resulting in lower
false positive rates for the same memory usage.

We performed six iterations of unique k-mer trimming on 5 million
Illumina reads from sequencing of {\em E. coli}, with memory
usage less than 30 MB.  For each iteration we measured empirical false
positive rate compared with number of bases trimmed as well as the
total number of k-mers (Table \ref{table:loop_trim}).  In the first
round, the estimated false positive rate was 80.0\%, and 13.5\% of the
total bases were removed by trimming reads at low-abundance k-mers;
the second iteration had a false positive rate of 37.7\%, and removed
only 1.5\% additional data; and by the fourth iteration the false
positive rate was down to 23.2\% with 0.0\% of the data removed.

The elimination of so many unique k-mers (column 5) in the first pass
was unexpected: the high false positive rate should have resulted in
fewer k-mers being identified as unique, were the erroneous
k-mers independent of each other. Upon examination, we realized that
in Illumina data erroneous k-mers typically come from substitution
errors that yield runs of up to k erroneous k-mers in a row
\cite{Kelley2010}.  When trimming reads with high false positive
rates, these runs are typically trimmed after the first few unique
k-mers, leaving unique k-mers at the 3' end.  Because of this we
hypothesized that high-FP rate trimming would result in the retention
of many unique k-mers at the 3' end of the read, and this was
confirmed upon measurement (Table \ref{table:loop_trim}, column 6,
pass 1 vs pass 2).

In comparison to quality-based trimming software such as seqtk and
FASTX, trimming at unique k-mers performed very well: in this data
set, all unique k-mers represent errors, and even with an initial
false positive rate of 80\%, khmer outperformed all but the most
stringent seqtk run (Table \ref{table:loop_trim}).  With a lower false
positive rate or multiple passes, khmer eliminates more erroneous
k-mers than seqtk or FASTX.  The tradeoff here is in memory usage:
for larger data sets, seqtk and FASTX will consume
the same amount of memory as on smaller data sets, while khmer's memory
usage will need to grow with the data set size.

\subsection*{Using khmer for digital normalization, a streaming algorithm}

Digital normalization is a lossy compression algorithm that discards
short reads based on saturating coverage of a de Bruijn graph
\cite{Brown2012}.  While several non-streaming implementations exist,
including Trinity's {\em in silico} normalization
\cite{Haas2013,Brown2012blog}, digital normalization can be
efficiently implemented as a {\em streaming} algorithm. In the
streaming implementation, if a read is not kept, it is not loaded into
the Count-Min Sketch structure, and the false positive rate does not
increase.  For high coverage data sets, the digital normalization
algorithm is sublinear in memory because it does not collect the
majority of k-mers in those data sets \cite{Brown2012}.  This has the
advantage of enabling low-memory preprocessing of both high-coverage
genomic data sets, as well as mRNAseq or metagenomic data sets with
high-coverage components \cite{Brown2012, Howe2012}.

While digital normalization is already implemented inside khmer,
previous work did not explore the lower bound on memory usage for
effective digital normalization.  In particular, the effects of high
false positive rates have not been examined in any prior work.

We applied digital normalization to the {\em E. coli} data set used
above, and chose seven different Count-Min Sketch sizes to yield seven
different false positive rates \ref{table:loop_norm}.  The data set
was normalized to a k-mer coverage of 20 and the resulting data were
evaluated for retention of true and erroneous k-mers, as in
\cite{Brown2012} (Table \ref{table:loop_norm}).  The results show that
digital normalization retains the same set of underlying ``true''
k-mers until the highest false positive rate of 100\% (Table
\ref{table:loop_norm}, column 5), while discarding only about 2\%
additional reads (Table \ref{table:loop_norm}, column 6).

To evaluate the effect of digital normalization with high false
positive rates on actual genome assembly, we next performed
normalization to a coverage of 20 with the same range of false
positive rates as above.  We then assembled this data with Velvet
\cite{Zerbino2008} and compared the resulting assemblies to the known
     {\em E. coli} MG1655 genome using QUAST (Table
     \ref{table:assembly}).  To our surprise, we found that even after
     executing digital normalization with a false positive rate of
     83.2\%, a nearly complete assembly was generated.  No progressive
     increase in misassemblies (measured against the real genome with
     QUAST) was seen across the different false positive rates (data
     not shown). This suggests that below 83.2\% FP rate, the false
     positive rate of digital normalization has little to no effect on
     assembly quality with Velvet.  (Note that the Velvet assembler
     itself used considerably more memory than digital normalization.)


While these results are specific to Velvet and the coverage parameters
used in digital normalization, they do suggest that no significant
information loss occurs due to false positive rates below 80\%.
Further evaluation of assembly quality in response to different
normalization parameters and assemblers is beyond the scope of of this
paper.

\section*{Discussion}

\subsection*{khmer enables fast, memory-efficient online counting}

khmer enables memory- and time-efficient online counting (Figures
\ref{fig:cmp_time}, \ref{fig:cmp_memory}, and \ref{fig:cmp_count}).
This is particularly important for the streaming approaches to data
analysis needed as data set sizes increase.  Because query and
updating of k-mer counts can be done directly as data is being loaded,
with no need for disk access or an indexing step, khmer can also
perform well in situations with poor disk I/O performance.  (Note that
BFCounter also supports online k-mer counting \cite{Melsted2011}.)

\subsection*{khmer is a generally useful k-mer counting approach}

In addition to online counting, khmer offers a general range of useful
performance tradeoffs for disk I/O, time and memory.  From the
performance comparison between khmer and other k-mer counting packages
in calculating k-mer abundance distributions, khmer is comparable with
existing packages.  In time, khmer performs competitively with DSK and
BFCounter (Figure \ref{fig:cmp_time}); khmer also provides a way to
systematically trade memory for miscounts across a wide range of
parameters (Figure \ref{fig:cmp_memory}).  khmer's uncompressed disk
storage is competitive with Jellyfish, and, in situations where disk
space is at a premium, khmer can take advantage of gzip compression to
provide storage similar to that of DSK (Figure \ref{fig:cmp_disk},
purple line with boxes).

KMC, DSK, and KAnalyze perform especially well in memory usage for
calculating the abundance distribution of k-mers. However, in exchange
for this efficiency, retrieving specific k-mer counts at random is
likely to be quite slow, as DSK is optimized for iterating across
partition sets of k-mers rather than randomly accessing k-mer counts.

For retrieving the counts of individual k-mers, khmer is significantly
faster than both Tallymer and Jellyfish.  This is not surprising,
since this was a primary motivation for the development of khmer.

\subsection*{khmer memory usage is fixed and low}

The memory usage of the basic Count-Min Sketch approach is fixed:
khmer's memory usage does not increase as data is loaded. While this
means that khmer will never crash due to memory limitations, and all
operations can be performed in main memory without recourse to disk
storage, the false positive rate may grow too high.  Therefore the
memory size must be chosen in light of the false positive rate and
miscount acceptable for a given application.  In practice, we
recommend choosing the maximum available memory, because the false
positive rate decreases with increasing memory and there are no
negative effects to minimizing the false positive rate.

For any given data set, the size and number of hash tables will
determine the accuracy of k-mer counting with khmer.  Thus, the user
can control the memory usage based on the desired level of accuracy
(Figure \ref{fig:cmp_memory}). The time usage for the first step of
k-mer counting, consuming the reads, depends on the total amount of
data, since we must traverse every k-mer in every read.  The second
step, k-mer retrieval, is algorithmically constant for fixed k;
however, for practicality, the hash tables are usually saved to and
loaded from disk, meaning that k-mer retrieval time depends directly
on the size of the database being queried.

The memory usage of khmer is particularly low for sparse data sets,
especially since only main memory is used and no disk space is
necessary beyond that required for the read data sets.  This is no
surprise: the information theoretic comparison in \cite{Pell2012}
shows that, for sparse sequencing data sets, Bloom filters require
considerably less memory than any possible exact information storage
for a wide range of false positive rates and data set sparseness.

In our implementation we use 1 byte to store the count of each k-mer
in the data structure. Thus the maximum count for a k-mer will be 255.
In cases where tracking bigger counts is required, khmer also provides
an option to use an STL map data structure to store counts above 255,
with the trade-off of significantly higher memory usage.  In the
future, we may extend khmer to counters of arbitrary bit sizes.


\subsection*{False positive rates in k-mer counting are low and predictable}

The Count-Min Sketch is a probabilistic data structure with a one-sided
error that results in random overestimates of k-mer frequency, but
does not generate underestimates.

In the Count-Min Sketch, the total memory usage is fixed; the memory
usage, the hash functions, and the total number of distinct objects
counted all influence the accuracy of the count.  While the
probability of an inaccurate count can easily be estimated based on
the hash table load, the miscount size is dependent on details of the
frequency distribution of k-mers \cite{Cormode2005}.

More specifically, in the analysis of the Count-Min Sketch, the
difference between the incorrect count and actual count is related to
the total number of k-mers in a data set and the size of each hash
table \cite{Cormode2005}. Further study has shown that the behavior of
Count-Min Sketch depends on specific characteristics of the data set
under consideration, especially left-skewness \cite{Rusu2008,
  CormodeM05}.  These probabilistic properties suit short reads from
next generation sequencing data sets: the miscounts are low because of
the highly left-skewed abundance distribution of k-mers in these data
sets.

Figures \ref{fig:average_offset_vs_fpr} and
\ref{fig:percent_offset_vs_fpr} demonstrate these properties well.  We
see more correct counting for error-prone reads from a genome than for
error-free reads from a genome, with a normal distribution of k-mer
abundance.  Thus, this counting approach is especially suitable for
high diversity data sets, such as metagenomic data, in which a larger
proportion of k-mers are low abundance or unique due to sequencing
errors.

\subsection*{Real-world applications for khmer}

For many applications, an approximate k-mer count is sufficient.  For
example, when eliminating reads with low abundance k-mers, we can
tolerate a certain number of low-frequency k-mers remaining in the
resulting data set falsely.  If RAM-limited we can do the filtering
iteratively so that at each step we are making more effective use of
the available memory.

In practice, we have found that a false positive rate of between 1\%
and 10\% offers acceptable miscount performance for a wide range of
tasks, including error profiling, digital normalization and
low-abundance read-trimming.  Somewhat surprisingly, false positive rates of up
to 80\% can still be used for both read trimming and digital
normalization in memory-limited circumstances, although multiple
passes across the data may be needed.

For many applications, the fact that khmer does not break an imposed
memory bound is extremely useful, since for many data sets ---
especially metagenomic data sets --- high memory demands constrain
analysis \cite{Howe2012,Luo2009}.  Moreover, because the false
positive rate is straightforward to measure, the user can be warned
that the results should be invalidated when too little memory is used.
When combined with the graceful degradation of performance for both
error trimming and digital normalization, khmer readily enables
analysis of extremely large and diverse data sets \cite{adina2013}. In
an experiment to assemble the reads of a soil metagenomic sample
collected from Iowa prairie, the number of reads to assemble drops
from 3.3 million to 2.2 million and the size of the data set drops from
245GB to 145GB accordingly after digital normalization
\cite{Howe2012}.  240GB memory was used in the process. This also
shows that khmer works well to analyze large, real-world metagenomic data
sets.

\subsection*{Conclusion}

K-mer counting is widely used in bioinformatics, and as sequencing
data set sizes increase, graceful degradation of data structures in
the face of large amounts of data has become important.  This is
especially true when the theoretical and practical effects of the
degradation can be predicted (see e.g. \cite{Melsted2011, Pell2012,
  Roy2014}).  This is a key property of the Count-Min Sketch approach,
and its implementation in khmer.

The khmer software implementation offers good performance, a robust
and well-tested Python API, and a number of useful and well-documented
scripts.  While Jellyfish, DSK, KMC, and Turtle also offer good
performance, khmer is competitive, and, because it provides a Python
API for online counting, is flexible.  In memory-limited situations
with poor I/O performance, khmer is particularly useful, because it
will not break an imposed memory bound and does not require disk
access to store or retrieve k-mer counts.  However, in exchange for
this memory guarantee, counting becomes increasingly incorrect as less
memory is used or as the data set size grows large; in many situations
this may be an acceptable tradeoff.

\subsection*{Future considerations}

Applying khmer to extremely large data sets with many distinct k-mers
requires a large amount of memory: approximately 446 GB of memory is
required to achieve an false positive rate of 1\% for $50\times 10^9$
k-mers. It is possible to reduce the required memory by dividing k-mer
space into multiple partitions and counting k-mers separately for each
partition. Partitioning k-mer space into $M$ partitions results in a
linear decrease in the number of k-mers under consideration, thus
reducing the occupancy by a constant factor $M$ and correspondingly
reducing the collision rate.  Partitioning k-mer space is a
generalization of the systematic prefix filtering approach, where one
might first count all k-mers starting with AA, then AC, then AG, AT,
CA, etc., which is equivalent to partitioning k-mer space into 16
equal-sized partitions. These partitions can be calculated
independently, either across multiple machines or iteratively on a
single machine, and the results stored for later comparison or
analysis.  This is similar to the approach taken by DSK
\cite{Rizk2013}, and could easily be implemented in khmer.

Further optimization of khmer on single machines, e.g. for multi-core
architectures, is unlikely to achieve significantly greater speed.
Past a certain point k-mer counting is fundamentally I/O bound
\cite{McDonald2013}.

Perhaps the most interesting future direction for probabilistic k-mer
counting is that taken by Turtle \cite{Roy2014}, in which several data
structures are provided, each with different tradeoffs, but with a
common API.  We hope to pursue this direction in the future by
integrating such approaches into khmer.

\section*{Methods}

\subsection*{Code and data set availability}


The version of khmer used to generate the results below is available
at http://github.com/ged-lab/khmer.git, tag '2013-khmer-counting'.
Scripts specific to this paper are available in the paper repository
at \\
https://github.com/ged-lab/2013-khmer-counting.
The IPython\cite{4160251} notebook file and data analysis to generate the figures are also
available in that github repository.  Complete instructions to reproduce
all of the results in this paper are available in the khmer-counting
repository; see README.rst.

\subsection*{Sequence Data}

One human gut metagenome reads data set (MH0001) from the 
MetaHIT (Metagenomics of the Human Intestinal Tract) project \cite{Qin2010} was used. 
It contains approximately 59 million reads, each 44bp long; it was trimmed to remove 
low quality sequences. 

Five soil metagenomics reads data sets with different size were taken
from the GPGC project for benchmark purpose (see Table
\ref{table:datasets}).  These reads are from soil in Iowa region and they
are filtered to make sure there are less than 30\% Ns in the read and
each read is longer than 30 bp.  The exact data sets used for the
paper are available on Amazon S3 and the instructions to acquire these
data sets are available in the paper repository on github.com.

We also generated four short-read data sets to assess the false
positive rate and miscount distribution. One is a subset of a real
metagenomics data set from the MH0001 data set, above. The second
consists of randomly generated reads. The third and fourth contain
reads simulated from a random, 1 Mbp long genome.  The third has a
substitution error rate of 3\%, and the fourth contains no errors. The
four data sets were chosen to contain identical numbers of distinct
22-mers.  The scripts necessary to regenerate these data are available
in the paper repository on github.com.

\subsection*{Count-Min Sketch implementation}

We implemented the Count-Min Sketch data structure, a simple
probabilistic data structure for counting distinct elements
\cite{Cormode2005}.  Our implementation uses $Z$ independent hash
tables, each containing a prime number of counters $H_i$.  The hashing
function for each hash table is fixed, and reversibly converts each
DNA k-mer (for $k \le 32$) into a 64-bit number to which the modulus of
the hash table size is applied.  This provides $Z$ distinct hash
functions.

To increment the count associated with a k-mer, the counter associated
with the hashed k-mer in each of the $N$ hash tables is incremented.
To retrieve the count associated with a k-mer, the minimum count
across all $N$ hash tables is chosen.

In this scheme, collisions are explicitly not handled, so the count
associated with a k-mer may not be accurate. Because collisions only
falsely {\em increment} counts, however, the retrieved count for any
given k-mer is guaranteed to be no less than the correct count.  Thus
the counting error is one-sided.

\subsection*{Hash function and khmer implementation}

The current khmer hash function works only for $k \le 32$ and converts
DNA strings exactly into 64-bit numbers.  However, any hash function
would work. For example, a cyclic hash would enable khmer to count
k-mers larger in size than 32; this would not change the scaling
behavior of khmer at all, and is a planned extension.

By default khmer counts k-mers in DNA, i.e. strandedness is
disregarded by having the hash function choose the lower numerical
value for the exact hash of both a k-mer and its reverse complement.
This behavior is configurable via a compile-time option.

\subsection*{Comparing with other k-mer counting programs}

We generated k-mer abundance distribution from five soil metagenomic reads
data sets of different sizes using khmer, Tallymer, Jellyfish, DSK, BFCounter, 
KMC, Turtle and KAnalyze. If the software 
does not include function to generate k-mer abundance distribution directly, we output the
frequency of each k-mer in an output file.
We fixed k at 22 unless otherwise noted.

\paragraph{khmer:}
For khmer, we set hash table sizes to fix the false positive rate at
either 1\%, 5\% or 20\%, and used 8 threads in loading the data.

We did the khmer random-access k-mer counting benchmark with a
custom-written Python script {\tt khmer-count-kmers} which loaded the
database file and then used the Python API to query each k-mer
individually.

\paragraph{Tallymer:}
Tallymer is from the genometools package version 1.3.4. For the {\tt suffixerator} subroutine 
we used:
{\tt -dna -pl -tis -suf -lcp}.

We use the {\tt mkindex} subroutine to generate k-mer abundance distribution, we used:
{\tt -mersize 22}.

The Tallymer random access k-mer counting benchmark was done using the
'tallymer search' routine against both strands; see the script
{\tt tallymer-search.sh}.

\paragraph{Jellyfish:}
The Jellyfish version used was 1.1.10 and the multithreading option is set to 8 threads.

Jellyfish uses a hash table to store the k-mers and the size of the
hash table can be modified by the user.  When the specified hash table
fills up, Jellyfish writes it to the hard
disk and initializes a new hash table.  Here we use a
similar strategy as in \cite{Melsted2011} and chose the minimum size of the hash 
tables for Jellyfish so that all k-mers were stored in memory.

We ran Jellyfish with the options as below:

{\tt jellyfish count -m 22 -c 2 -C} for k=22.


The Jellyfish random access k-mer counting benchmark was performed
using the 'query' routine and querying against both strands; see
the script {\tt jelly-search.sh}.

\paragraph{DSK:} We ran DSK with default parameters with {\tt -histo} option to generate
k-mer abundance distribution. The DSK version used was 1.5031.

\paragraph{BFCounter:} The BFcounter version used was 1.0 and the multithreading option is set to 8
 threads.
 
We ran BFCounter {\tt count} subroutine with the options as below:
 
 {\tt BFCounter count -k 22 -t 8 -c 100000}.
 {\tt -n} option representing the estimated number of k-mers is adjusted to the different 
 test data sets.
 
 This subroutine produces the actual count of k-mers in input files.
 
We ran BFCounter {\tt dump} subroutine with the options as below:
 {\tt BFCounter dump -k 22}.
 
This subroutine can write k-mer occurrences into a tab-separated text file.

\paragraph{KMC:} The KMC version used was 0.3. We ran both {\tt kmc} and {\tt kmc\_dump} subroutines 
with default parameters.

\paragraph{Turtle:} The Turtle version used was 0.3. We ran {\tt scTurtle32} with the multithreading 
option set to 8 threads and {\tt -n} option representing expected number of frequent k-mers
is adjusted to different test data sets.

\paragraph{KAnalyze:} The KAnalyze version used was 0.9.3. We ran {\tt count} subroutine 
with default parameters.


\section*{Acknowledgments}

We thank Eric McDonald for technical assistance with optimizing the khmer codebase.

%

\bibliography{khmer-counting}
\clearpage

\section*{Figure Legends}



\clearpage
\section*{Tables}

\begin{table}[!ht]
\caption{
\bf{Benchmark soil metagenome data sets for k-mer counting performance, taken from
\cite{Howe2012}.}}
\begin{tabular}{ |c | c |c| c|c| }
\hline
Data set & size of file (GB) & number of reads & number of distinct
k-mers & total number of k-mers \\
\hline
subset 1        & 1.90 &  9,744,399 &   561,178,082 &   630,207,985 \\
subset 2        & 2.17 & 19,488,798 & 1,060,354,144 & 1,259,079,821 \\
subset 3        & 3.14 & 29,233,197 & 1,445,923,389 & 1,771,614,378 \\
subset 4        & 4.05 & 38,977,596 & 1,770,589,216 & 2,227,756,662 \\
entire data set & 5.00 & 48,721,995 & 2,121,474,237 & 2,743,130,683 \\
\end{tabular}
\begin{flushleft}
\end{flushleft}
\label{table:datasets}
\end{table}



%

\begin{table}[!ht]
\caption{
\bf{Data sets used for analyzing miscounts.}}
\begin{tabular}{ | p{5cm} | c | c | c |}
\hline
Data set & Size of data set file & Number of total k-mers & Number of distinct k-mers \\
\hline
Real metagenomics reads                                  & 7.01M  & 2,917,200  & 1,944,996 \\
\hline
Totally random reads with randomly generated k-mers      & 3.53M  & 2,250,006  & 1,973,059 \\
\hline
Simulated reads from simulated genome with error         & 5.92M  & 3,757,479  & 2,133,592 \\
\hline
Simulated reads from simulated genome without error      & 9.07M  & 5,714,973  & 1,989,644 \\
\hline
Real {\em E. coli} reads                                        & 4.85M  & 4,004,911  & 2,079,302 \\
\end{tabular}
\begin{flushleft}
\end{flushleft}
\label{table:random_data}
\end{table}


\begin{table}[!ht]
\caption{
\bf{Iterative low-memory k-mer trimming.  The results of trimming
  reads at unique (erroneous) k-mers from a 5m read {\em E. coli} data set (1.4 GB)
  in under 30 MB of RAM.  After each iteration, we measured the
  total number of distinct k-mers in the data set, the total number
  of unique (and likely erroneous) k-mers remaining, and the
  number of unique k-mers present at the 3' end of reads.}}
\begin{tabular}{ | c | c | c | c | c | c |}
\hline
 & FP rate & bases trimmed & distinct k-mers & unique k-mers & 
unique k-mers at 3' end \\
\hline
untrimmed                           &      -  &      - & 41.6m & 34.1m & 30.4\%  \\
khmer iteration 1                   & 80.0\%  & 13.5\% & 13.3m &  6.5m & 29.8\% \\
khmer iteration 2                   & 40.2\%  &  1.7\% &  7.6m & 909.9k & 12.3\% \\
khmer iteration 3                   & 25.4\%  &  0.3\% &  6.8m & 168.1k & 3.1\% \\
khmer iteration 4                   & 23.2\%  &  0.1\% &  6.7m &  35.8k & 0.7\% \\
khmer iteration 5                   & 22.8\%  &  0.0\% &  6.6m &   7.9k & 0.2\% \\
khmer iteration 6                   & 22.7\%  &  0.0\% &  6.6m &   1.9k & 0.0\% \\
filter by FASTX                     &      -  &  9.1\% & 26.6m & 20.3m & 26.3\% \\
filter by seqtk(default)            &      -  &  8.9\% & 17.7m & 12.1m & 12.3\% \\
filter by seqtk(-q 0.01)            &      -  & 15.4\% &  9.9m &  5.1m &  5.2\% \\
filter by seqtk(-b 3 -e 5)          &      -  &  8.0\% & 34.5m & 27.7m & 25.3\% \\
\end{tabular}
\begin{flushleft}
\end{flushleft}
\label{table:loop_trim}
\end{table}


\begin{table}[!ht]
\caption{ \bf{Low-memory digital normalization. The results of
    digitally normalizing a 5m read {\em E. coli} data set (1.4 GB) to C=20
    with k=20 under several memory usage/false positive rates.  The
    false positive rate (column 1) is empirically determined.  We
    measured reads remaining, number of ``true'' k-mers missing from
    the data at each step, and the number of total k-mers remaining.
    Note: at high false positive rates, reads are erroneously removed due to
    inflation of k-mer counts.}}
\begin{tabular}{ | c | c | c | c | c | c | c |}
\hline
memory   & FP rate & retained reads & retained reads \% & true k-mers missing & total k-mers \\
\hline
before diginorm   &  -      & 5,000,000   & 100.0\%    & 170  &  41.6m \\
2400 MB           &   0.0\% & 1,656,518   &  33.0\%    & 172  &  28.1m \\
240 MB            &   2.8\% & 1,655,988   &  33.0\%    & 172  &  28.1m \\
120 MB            &  18.0\% & 1,652,273   &  33.0\%    & 172  &  28.1m \\
60 MB             &  59.1\% & 1,633,182   &  32.0\%    & 172  &  27.9m \\
40 MB             &  83.2\% & 1,602,437   &  32.0\%    & 172  &  27.6m \\
20 MB             &  98.8\% & 1,460,936   &  29.0\%    & 172  &  25.7m \\
10 MB             & 100.0\% & 1,076,958   &  21.0\%    & 185  &  20.9m \\
\end{tabular}
\begin{flushleft}
\end{flushleft}
\label{table:loop_norm}
\end{table}


\begin{table}[!ht]
\caption{
\bf{{\em E. coli} genome assembly after low-memory digital normalization.
  A comparison of assembling reads digitally normalized with low memory/high
  false positive rates.  The reads were digitally normalized to 
  C=20 (see
  \cite{Brown2012} for more information) and were assembled using Velvet.
  We measured total length of assembly,
  as well as percent of true MG1655 genome covered by the assembly using QUAST.}}
\begin{tabular}{ | c | c | c | c | c | c |}
\hline
memory   & FP rate & N contigs & total length(bases) & \% of true genome covered \\
\hline
before diginorm  &-   & 106 & 4,546,051 & 97.84\% \\
    2400 MB  &  0.0\% & 617 & 4,549,235 & 98.05\% \\
     240 MB  &  2.8\% &  87 & 4,549,253 & 98.04\% \\
     120 MB  & 18.0\% &  86 & 4,549,335 & 98.04\% \\
      60 MB  & 59.1\% &  90 & 4,548,619 & 98.03\% \\
      40 MB  & 83.2\% &  89 & 4,550,599 & 98.11\% \\
      20 MB  & 98.8\% &  85 & 4,550,014 & 98.04\% \\
      10 MB  &100.0\% &  97 & 4,545,871 & 97.97\% \\
\end{tabular}
\begin{flushleft}
\end{flushleft}
\label{table:assembly}
\end{table}

\end{document}